\documentstyle[12pt,aasms4]{article}
\begin{document}

\title{Do Statistically Significant Correlations Exist between\\
the Homestake Solar Neutrino Data and Sunspots?}

\author{J. Boger, R.L. Hahn, and J.B. Cumming}
\affil{Chemistry Department, Brookhaven National Laboratory, Upton,
New York 11973-5000}
\authoremail{cumming@nc8.chm.bnl.gov}

\begin{abstract}
It has been suggested by various authors that a significant anticorrelation
exists between the Homestake solar neutrino data and the sunspot cycle.
Some of these claims rest on smoothing the data by taking running averages,
a method that has recently undergone criticism. We demonstrate that
no significant anticorrelation can be found in the Homestake data,
or in standard 2- and 4-point averages of that data. However, when 3-, 5-,
and 7-point running averages are taken, an anticorrelation seems
to emerge whose significance grows as the number of points
in the average increases. Our analysis indicates that the apparently high
significance of these anticorrelations is an artifact of the failure
to consider the loss of independence introduced in the running average
process. When this is considered, the significance is reduced to that of the
unaveraged data. Furthermore, when evaluated via parametric
subsampling, no statistically significant anticorrelation is found.
We conclude that the Homestake data can not be used to substantiate
any claim of an anticorrelation with the sunspot cycle.
\end{abstract}

\keywords{neutrinos, sunspots, correlations}

\section{Introduction}

The chlorine solar-neutrino detector, located in the Homestake gold mine
in Lead, South Dakota, was the first to successfully detect
neutrinos from the Sun and is the longest running solar neutrino
experiment to date. The rate of neutrino captures was soon
observed (Rowley, Cleveland, \& Davis 1984) to be much lower than that
predicted by the Standard Solar Model (SSM)(Bahcall 1989).
SSM calculations (Turck-Chi\'eze \& Lopes 1993,
Bahcall, Basu, \& Pinsonneault 1998)
predict capture rates between 1.4 and
1.7 atoms/d for the chlorine experiment.
The most recent experimental value (Cleveland et al. 1998) 
is only $0.479 \pm 0.043$ (combined statistical and systematic errors)
atoms/d, or 27-34\% of the expected signal.
This deficit, generally referred to as the ``solar neutrino problem,''
has now been observed by other experiments (Hirata et al. 1998,
Fukuda et al. 1998, Abdurashitov et al. 1999, Hampel et al. 1999).
A number of explanations have been considered. Some of these involve
modifications of the SSM; others propose new physics beyond the Standard Model
of particle physics, see the summary by Bahcall (1989).
With regard to the Homestake data, it has been suggested
that the solar neutrino signal varies with time, and further, that
the signal is anticorrelated with the well-known sunspot
cycle (Rowley et al. 1984, Davis 1986,
Davis, Cleveland, \& Rowley 1987) 
or with other indicators of solar activity 
( Massetti, Storini, \& Iucci 1995).

To explain this anticorrelation, it has been proposed that 
the neutrino has a magnetic moment. However, such a claim rests on 
the significance of the anticorrelation. Various authors have examined 
the Homestake data, some finding little or no evidence for any 
anticorrelation, and others finding a significantly large one.
Of those who find significant anticorrelations, details of the
analysis are often not given. In some cases,
the data are smoothed by taking running averages. However,
Walther (1997, 1999) has argued that the method of running 
averages may mistakenly
lead to significant anticorrelations. To illustrate this, Walther (1997)
generated a set of {\it x\/}-values by randomly selecting them from a 
normal distribution. He did the same to generate a set of {\it y\/}-values but
from a different normal distribution. These points are inherently 
uncorrelated. He then took a ten-point running average of this data set
and found the apparent significance of the correlation to increase
when compared with the non-smoothed data. In a subsequent paper,
Walther (1999) employed a statistical method known as parametric
subsampling, a procedure to evaluate data when the points are not
independent, to assess the significance of the Homestake data when
smoothed with running averages. He found no significant anticorrelation
between the Homestake data and sunspots.

Following this lead, we undertook a detailed reexamination of the 
Homestake and sunspot data. In
Sec.$\>2$, below, some details of the Homestake experiment are reviewed
and its basic results, $^{37}\!$Ar production rates as a function of 
time, are examined. We did not find significant variations
anticorrelated with sunspot numbers. The method of weighted running
averages is applied in Sec.$\>3$ to smooth the Homestake and sunspot data.
An apparent anticorrelation emerges whose significance increases with the
number of points used to make the averages. Our statistical analysis of the
Homestake data when smoothed with running averages differs from Walther's
but arrives at the same conclusion of no significant anticorrelation; our
correlation coefficients have significances similar to the unsmoothed data.
This suggests that the noted increase in significance is an artifact related
to the failure to consider the reduction in the number of independent
points when running averages are taken. We arrive at the same conclusion when, 
like Walther, we apply parametric subsampling to the analysis of the 
Homestake data. We conclude that, when analyzed properly, no significant 
anticorrelation exists between the Homestake solar neutrino data and 
sunspot numbers.
       
\section{The Homestake Experiment: Basic Results}

The Homestake detector contains 615 tons of
perchlorethylene (C$_2$Cl$_4$) located 1478 m underground
in a gold mine.
It utilizes a radiochemical procedure based on the inverse
$\beta$-decay reaction
$$  \nu_e + ^{37}\!{\rm Cl} \rightarrow ^{37}\!\!\!{\rm Ar} + e^-.  \eqno{(1)}  $$
The $^{37}\!$Ar produced by neutrino capture on stable $^{37}$Cl
decays by electron capture back to $^{37}$Cl with
a half life of 35.0 d. Reaction (1) has a threshold of 814 keV, hence
Homestake is sensitive only to electron neutrinos from the {\it pep} reaction
and those emitted by the decay of
$^7$Be and $^8$B in the Sun. It is transparent to other neutrino flavors.

At the start of a ``run,'' stable Ar carrier gas is
added to the tank. After $\sim\!3$ months,
the carrier Ar and any $^{37}\!$Ar produced during the exposure are removed
from the detector by sweeping with He gas.
The recovered Ar (yields $\sim\!95$\%) is purified by gas chromatography
and transferred into a proportional counter. The sample from each run
is assayed for approximately twelve $^{37}\!$Ar half lives. A maximum
likelihood fit (Cleveland 1983) is used to resolve the $^{37}\!$Ar decay
from the counter background.

The best-fit $^{37}\!$Ar production rates listed by Cleveland et al. (1998)
are plotted as a function of time in Fig. 1(a).
The measurements were nearly continuous from 1970.281 to 1994.388
with the exception of an $\sim\!1.4$-y gap (crosshatched in the figure)
due to the failure of two circulation pumps. Representative $68\%$
confidence ranges from the published work are shown as error bars in the
figure. This range is not, in general, symmetrically positioned about
the data point. We conservatively adopt the larger of the upper or lower
errors in the calculations which follow. With this weighting, the mean 
$^{37}\!$Ar production rate for the 108 runs is found to be $0.354\pm 0.028$
atoms/day. This is appreciably lower than the $0.479\pm 0.030$
(statistical only)atoms/day reported by Cleveland et al. (1998). Their value
was calculated via a maximum likelihood method which combined all the runs
into a single data set---essentially one $^{37}\!$Ar decay curve---whereas
for the purposes of our correlation analysis, we must keep each run discrete.
The difference between weighted mean and maximum likelihood value can be
traced to the fact that the lower production rates tend to
have smaller {\it absolute} errors than the higher ones, hence
are given more weight in the averaging process.
The unweighted mean is $0.485\pm 0.031$.

The mean sunspot number associated with each $^{37}\!$Ar measurement is plotted
as a function of time in Fig. 1(b). These are averages of daily sunspot
numbers (NOAA 1999) over the duration of the corresponding Homestake run.
The periodic behavior of the sunspots is apparent.
The $^{37}\!$Ar measurements commence at a time of decreasing
solar activity. They encompass two solar minima and maxima, and they end
near a third minimum. With this range, the data afford the possibility of
exploring correlations between the solar neutrino signal and sunspot number.

The dependence of $^{37}\!$Ar production rate on sunspot
number is shown in the form of a scatter plot in Fig. 1(c).
There is no obvious correlation or anticorrelation:
high and low rates are associated with both high and low sunspot numbers.
To quantify this, we define temporal regions in Fig. 1(a)
corresponding to three solar states: (1) when the Sun is quiet, at or near
the sunspot minima; (2) when the Sun is active,
at or near the sunspot maxima;
and (3) when the Sun is in transition between (1) and (2).
Weighted mean $^{37}\!$Ar production rates for each solar state
are listed in Table 1 together with the mean sunspot number for that
state. Were an anticorrelation to exist, one would expect the 
solar neutrino signal to be significantly larger when the Sun is quiet
rather than when it is active. Although the weighted mean production rate is 
slightly lower for the active Sun than for quiet conditions,
the $(10 \pm 19)\%$ reduction associated with a ten-fold increase in
mean sunspot number is clearly marginal. This conclusion is 
insensitive to the choice of weighting: the corresponding change
in unweighted means is $(16 \pm 15)\%$.
Mean production rates for these three solar states are the same at the
one-$\sigma$ level as that for all 108 measurements.

The correlation between $^{37}\!$Ar production and sunspot number was further
quantified by calculating (Press et al. 1992) Pearson's product-moment
coefficient, $r_p$, and Spearman's rank-order coefficient, $r_s$.
As can be seen in Table 2, values of $r_p$ and $r_s$
for the full 108 data points are comparable.
The significance of the weak anticorrelation, $r \sim\!-0.1$,
is normally considered in terms of the probability that the null hypothesis
is true, i.e., that an observed $r$ represents a statistical fluctuation
of otherwise uncorrelated data. The distribution of either $r_p$ or $r_s$
for $N$ independent but uncorrelated samples, the ``null distribution,'' 
is expected (Press et al. 1992) to be approximately normal
with a mean of zero and a standard deviation $\sigma = 1/{\sqrt{N-2}}$.
For $N = 108$, $\sigma = 0.097$. The one-sided probability that an observed
$r$ represents a statistical fluctuation, the ``null probability'' $P(r)$
in Table 2, is obtained by integrating the null distribution from $-1$ to $r$. 
There is a substantial probability, $\sim\!16 \%$, that the observed
anticorrelation is insignificant. (By convention, a correlation is
considered significant when the probability is less than $1\%$.) 
There is an equal probability for an accidental positive correlation.

The interpretation of Pearson's correlation coefficient is somewhat
dependent on the assumption that the distribution of the quantities
of interest is bivariate normal (Press et al. 1992).
The distribution of the Homestake $^{37}\!$Ar
production rates is shown as a histogram in Fig. 2(a).
The peak at or near zero reflects the fact that production rates
and lower $68\%$ confidence limits reported by Cleveland et al.(1998) 
were constrained to be $\ge 0$. Since the total number of events ($^{37}\!$Ar
plus background) recorded in an individual run is limited,  statistical
fluctuations in their temporal distribution might be expected to result in 
occasional negative values. The remaining data are consistent with the normal
distribution shown as a smooth curve in Fig. 2(a).
It peaks at $0.526\pm 0.042$, somewhat above the global maximum likelihood
value for the entire data set.
Sampling of a periodic function such as the sunspots
does not yield a normal distribution as is seen in Fig. 2(b).
We use Spearman's correlation coefficients in the following discussions as
they are less dependent on the assumption of a bivariate normal distribution. 
 
Values of $r_s$ and $P(r_s)$ for conventional 2- and 4-point averages of
the Homestake and sunspot data are also included in Table 2.
Such averaging procedures do not appear to improve
the significance of the anticorrelation. In fact, broadening of the
null distribution with decreasing number of points, to $\sigma = 0.200$
for $N=27$, leads to a reduced significance with a $27\%$ chance that the
observed $r_s = -0.125$ is due to a statistical fluctuation. While standard
averaging procedures may aid in the display of data, they entail
an intrinsic loss of information.

\section{Running Averages}

The method of running averages has been used by a number of
authors (Massetti et al. 1995, McNutt 1997) to smooth the Homestake
data, and claims have been made that $^{37}\!$Ar production rates
are anticorrelated with sunspot numbers or other
indicators of solar activity. While the present discussion is limited to
sunspot numbers, it has more general implications for
the use of running average procedures.
 
The weighted
running average, $X_i$, for run $i$ is defined as 
$$   X_i = \sum_{j=i-n}^{i+n}w_jx_j\!\bigg/\,\sum_{j=i-n}^{i+n}w_j,
\eqno{(2)} $$
and its weight, $W_i$, as 
$$   W_i = \sum_{j=i-n}^{i+n}w_j = \sum_{j=i-n}^{i+n}(1/\sigma_j^2). 
\eqno{(3)} $$
Here $x_j$, $\sigma_i$, and $w_j$ are the value, standard deviation,
and weight of the $j^{th}$ observed point.
The length of the running average is $l_a = 2n + 1$. As defined above,
$X_i$ corresponds to the midpoint of the $l_a$ points used in forming the
average.

Equations 2 and 3 are not applicable in the vicinity of an
``endpoint'' of which there are four in the Homestake data;
one at the beginning of the experiment,
a second at the start of the interruption due to the pump failure,
a third at the end of that failure,
and the last at the end of the measurements. That is, we did not average
across the gap due to the pump failure.
Following the method of Davis (1999), we define those
points directly adjacent to an endpoint as the unaveraged
value and weight, e.g., $X_1 = x_1$ and $W_1 = w_1$.
Those one removed from an
endpoint are defined as the three-point running averages, for example, 
$X_2$ is the weighted average of $x_1$, $x_2$, and $x_3$, and so on.

Three-, five-, and seven-point weighted running averages of the 108 points
that comprise the Homestake data set are plotted together with
the unaveraged ($l_a = 1$) values in Fig. 3. The period of data
interruption due to the pump failure is again crosshatched.
Some time dependent structure appears to emerge as the
length of the average increases. There is a hint of maxima (minima) for the
years 1977 (1980) and 1987 (1993), years that correspond to minima
(maxima) in the sunspot cycle (see Fig. 1(b)).    
Particularly apparent is the minimum in $^{37}\!$Ar
production at 1980, a time of maximum solar activity.
In their analysis of the first 61 Homestake runs,
Bahcall, Field \& Press (1987) ``conclude that the suggestive correlation
[Rowley et al. (1984)] between neutrino capture rate and sunspot number
depends almost entirely upon the four low points near the beginning of 1980.''
It remains an important feature of the 108-point data set now available. 
Note also that the high points on either side of the
pump failure, which persist as the length of the running average increases,
occur at a solar minimum.

Spearman's correlation coefficients presented in Table 3 suggest that the
use of running averages may reveal otherwise hidden correlations. 
The value of $r_s$ decreases from $-0.105$ for the unaveraged
data to $-0.240$ for $l_a = 7,$ and the apparent probability,
$P(r_s)$, that the null hypothesis is true $decreases$ from $14\%$
to $0.6\%$. We ask, is this valid evidence for an anticorrelation between 
the solar neutrino signal and sunspot number?

In calculating $P(r_s)$ it is assumed that the standard deviation of the null
distribution is determined by the original number of data points, $N = 108,$
such that $\sigma = 1/{\sqrt{106}}$. While standard
averaging reduces the number of points, running averaging
seems to maintain the full number. No information appears to be lost as
the procedure can be reversed to recover the original values and weights. 
However, the running-averaged points are not independent. If
we assume equal weights, it can be seen from Eq. 2 that the running average
for the $i^{th}$ run, $X_i$, is $(l_a - 1)\over l_a$ dependent on the
neighboring points. Alternatively, a measured value, $x_i$, contributes to
$l_a$ averaged points. The number of independent points is then on
the order of $N/l_a$. A better approximation
would include the special treatment of the four endpoints.
For example, the number of independent points for $l_a = 7$ can be
estimated to be
${4\over {1}} + {4\over {3}} + {4\over {5}} + {(108 - 12)\over {7}} = 19.8$. 
%pick whichever looks best!
%$4(1 + {1\over {3}} + {1\over {5}}) + (108 - 12){1\over {7}} = 19.8$. 
The predicted standard deviation of the null distribution is then
$\sigma = 1/{\sqrt{19.8-2}} = 0.237$, as compared to $\sigma = 0.097$
for the 108 independent data points. Values of $\sigma$ from this simple
estimation procedure are shown as a function of $l_a$ by
the smooth curve in Fig. 4. 

A null distribution can be generated by a Monte Carlo procedure which
first randomly shuffles the unaveraged $^{37}\!$Ar data. This breaks
any correlation with sunspots. One then performs the appropriate running
average on the shuffled data and calculates
Spearman's correlation coefficient between the averaged $^{37}\!$Ar values
and averaged sunspot numbers. Distributions were  
obtained from $10^6$ shuffles for $l_a = 1,3,5,$ and $7$.    
The calculated distribution for $l_a=1$, the unaveraged data, is
plotted as a histogram in Fig. 5. This is well fitted by a Gaussian,
the smooth curve in the figure. The standard deviation, $\sigma = 0.097$,
agrees with the value expected for $N = 108$ independent points.
The distribution for the 7-point running average in Fig. 5
is substantially broader.
The $\sigma = 0.230$ is that expected for 20.9 independent points.
There are visible deviations from the Gaussian fit.

Values of standard deviations obtained by this Monte Carlo procedure
for $l_a = 1,3,5$, and $7$ are shown as points in Fig. 4. While those
for $l_a = 3,5$, and $7$ fall somewhat below the curve described above, 
the large broadening introduced by the running averaging process
is confirmed. This broadening is not dependent on the
weighting of the average, e.g., the standard deviation of the null distribution
for the 7-point running average is 0.236 when weighting is applied compared
with 0.231 when it is not. Note that the broadening would not have been
observed if the $^{37}\!$Ar data had been shuffled {\it after}, rather than
before, the averaging process. 

The ``true'' null probabilities for each $r_s$
value in Table 3 were obtained by integration of the null distributions.  
The apparent increase in significance of the anticorrelation vanishes.
The reason for this can be seen in Fig. 5. The vertical line
at $r_s = -0.240$ indicates the correlation obtained for the 7-point
running average (Table 3). The area to the left of that line is
many-times larger for the true null distribution for $l_a = 7$ than
for that assuming 108 independent points. 

We have reached the same
conclusion as to lack of significance by applying Walther's parametric
subsampling procedure; the resulting probabilities are listed in
column 5 of Table 3. The reader is referred to Walther (1999) for details
of the calculation.

\section{Conclusions}

Claims have been made by some authors that the $^{37}\!$Ar production
rates measured in the Homestake solar neutrino experiment are anticorrelated
with sunspot number. Some of these rest on the use of running averages 
to smooth the data. It has been suggested by Walther(1997, 1999) 
that such a procedure
may lead to a substantial overestimation of the significance of the  
anticorrelations. We have critically reexamined 
the Homestake data, how the running averages have been applied, and
how the results have been interpreted. We reach the following conclusions:

1) Significant anticorrelations (at greater than an approximately
one-$\sigma$ level)
with the sunspot cycle are not found in the original Homestake data.

2) When the data are smoothed by taking running averages,
a significant anticorrelation does seem to emerge. Its apparent significance  
appears to increase with the number of points used to form the averages.

3) An analysis in terms of the true null distributions, as calculated by
Monte Carlo procedures, shows that the apparently high significance
of the anticorrelations for the 5- and 7-point running averages is an
artifact arising from the failure to consider the loss of independence
between neighboring points introduced by that averaging process. This
conclusion agrees with that inferred by the parametric subsampling procedure.

4) While the Homestake experiment has had a major impact on physics
by being the first to detect solar neutrinos and to identify
the solar neutrino problem, its present precision is
insufficient to substantiate the existence of a significant temporal
correlation with the sunspot cycle.

\acknowledgments
This work was supported by the Office of High Energy and Nuclear Physics of
the U.S. Department of Energy under contract No. DE-AC02-98CH10886.
We wish to thank R. Davis and J. Weneser for helpful discussions. 
%
% Table 1. SUN_TABLE.TEX -- $^{37}\!$Ar means for various solar conditions.
%
\clearpage
\begin{deluxetable}{cccc}
\tablenum{1}
\tablewidth{33pc}
%\tablewidth{0pc}
\tablecaption{Mean $^{37}\!$Ar production rates in atoms/d for various levels\protect\\
              of solar activity as measured by sunspot numbers.\label{tbl-1}}
\tablehead{
\colhead{Solar}          &\colhead{Number}          &
\colhead{Mean number}    &\colhead{Mean $^{37}\!$Ar}  \\
\colhead{status}         &\colhead{of runs}         &
\colhead{of sunspots}    &\colhead{production rate} }      
\startdata
    Active    & $  39 $ & $ 146 $ & $0.350\pm 0.047 $ \nl
     Quiet    & $  27 $ & $  15 $ & $0.389\pm 0.066 $ \nl
  Transition  & $  42 $ & $  59 $ & $0.341\pm 0.041 $ \nl
\enddata
\end{deluxetable}
%
% Table 2. RS_RP_TABLE.TEX -- Correlation Coefficients.
\begin{deluxetable}{cccc}
\tablenum{2}
\tablewidth{33pc}
%\tablewidth{0pc}
\tablecaption{Pearson's and Spearman's correlation coefficients\protect\\
              and null probabilities for the unsmoothed data\protect\\
              and 2- and 4-point averages.\label{tbl-2}}
\tablehead{
\colhead{Number}            &\colhead{Type of}          &
\colhead{Coefficient}       &\colhead{Null probability}  \\
\colhead{of points}         &\colhead{correlation\tablenotemark{a}}       &
\colhead{$r_p$ or $r_s$}    &\colhead{$P(r_p)$ or $P(r_s)$ in $\%$} }      
\startdata
     108    &   $ p $   &   $ -0.092 $   &   $ 17 $ \nl
     108    &   $ s $   &   $ -0.105 $   &   $ 14 $ \nl
      54    &   $ s $   &   $ -0.136 $   &   $ 16 $ \nl
      27    &   $ s $   &   $ -0.125 $   &   $ 27 $ \nl
\enddata
\tablenotetext{a}{$p =$ Pearson's, $s =$ Spearman's.} 
\end{deluxetable}
%
% Table 3. RS_TABLE.TEX --
%                 Rs values and estimates of their probabilities in %.
%
%\clearpage
\begin{deluxetable}{ccccc}
\tablenum{3}
\tablewidth{33pc}
%\tablewidth{0pc}
\tablecaption{Spearman's $r_s$ values and estimates of the probability\protect\\
              for the null hypothesis as a function of the\protect\\
              length of the running average. \label{tbl-3}}
\tablehead{
\colhead{Length}     & \colhead{Spearman's}     &
\colhead{$P(r_s)$ in $\%$}   & \colhead{$P(r_s)$ in $\%$}  &
\colhead{$P(r_s)$ in $\%$}   \\
\colhead{of Average, $l_a$} & \colhead{$r_s$} &
\colhead{Apparent}       & \colhead{True\tablenotemark{a}} &
\colhead{Subsampling\tablenotemark{b}}}      
\startdata
 1 & $-0.105$ & $14.0$ & $14.4$ & $15.5$ \nl
 3 & $-0.166$ & $ 4.3$ & $16.0$ & $16.0$ \nl
 5 & $-0.245$ & $ 0.6$ & $12.5$ & $15.0$ \nl
 7 & $-0.240$ & $ 0.6$ & $17.3$ & $20.0$ \nl
\enddata
\tablenotetext{a}{From null distributions estimated by the Monte Carlo
procedure; see text.}
\tablenotetext{b}{For a description of the method of parametric subsampling,
see Walther(1999).}
\end{deluxetable}
\clearpage

{\bf FIGURE CAPTIONS} 

Fig. 1. (a) $^{37}\!$Ar production rates (atoms/day) measured in the
Homestake experiment(Cleveland et al. 1998) as a function of time.
Error bars indicate $68\%$ confidence levels for some representative points.
(b) Mean sunspot numbers associated with each
$^{37}\!$Ar measurement, also as a function of time. Crosshatched areas
indicate the data interruption due to pump failure.
(c) $^{37}\!$Ar production rate as a function of the associated sunspot number.

Fig. 2. (a) The distribution of $^{37}\!$Ar production rates. The smooth curve
suggests an approximately normal distribution (see text).
(b) The distribution of associated mean sunspot numbers. 

Fig. 3. Growth of apparent temporal structure on applying weighted
running averages to the measured $^{37}\!$Ar production rates.
(a) The original data. (b) Three point running averages. (c) Five
point running averages. (d) Seven point running averages.  
Crosshatched areas indicate the data interruption due to pump failure.

Fig. 4. Standard deviation of the null distribution for
Spearman's correlation coefficient as a function of
the length of the running average. The curve shows the broadening suggested
by a simple estimation procedure; see text.
The points are from the null distributions generated by the Monte Carlo
random shuffling of the original $^{37}\!$Ar data.

Fig. 5. Null distributions of Spearman's correlation coefficient
between $^{37}\!$Ar production rates and sunspot numbers for the original
Homestake data and for those data subjected to seven-point running averaging.  
Histograms show the results generated by a random shuffling procedure;
see text. A Gaussian fit to each histogram is shown as a smooth curve.  
The vertical line indicates the $r_s = -0.24$ obtained for the unshuffled
7-point running average of the data.

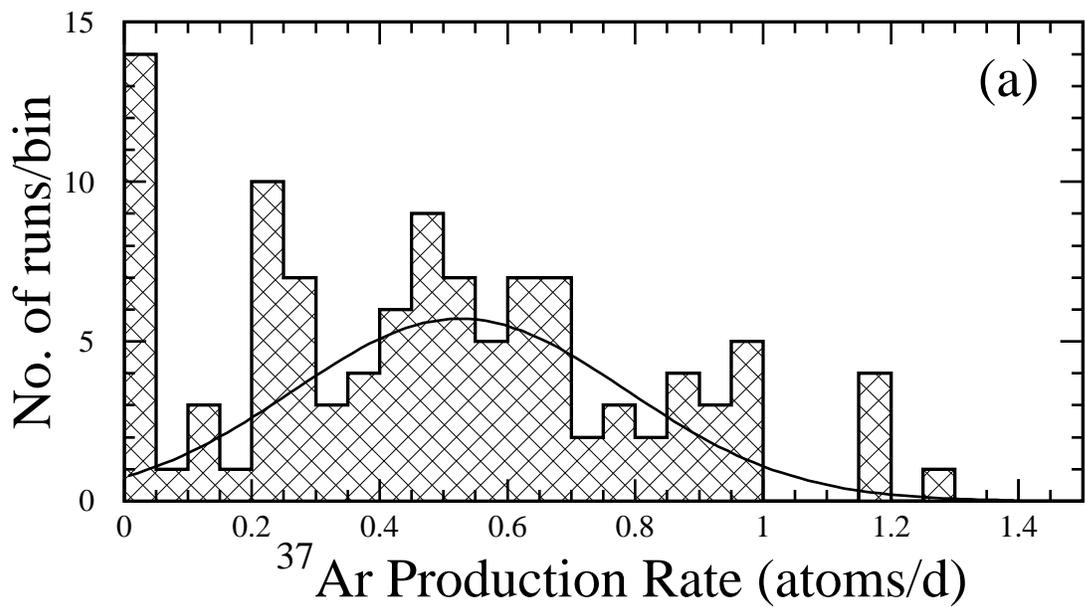
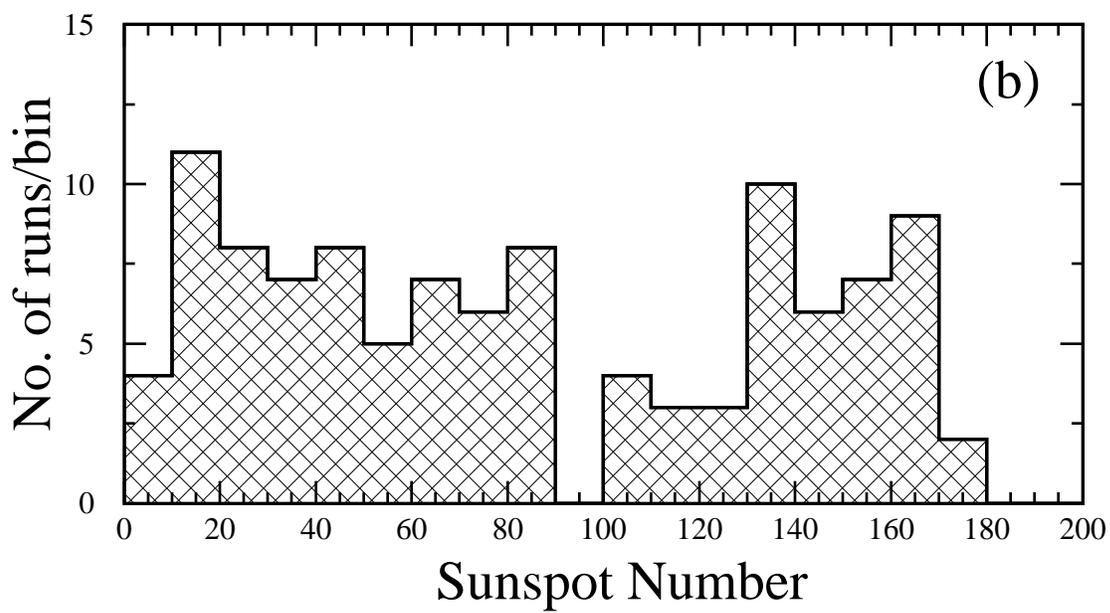
\end{document}